\documentclass[aps,eqsecnum,showpacs,floatfix,preprint]{revtex4}

\usepackage{dcolumn}
\usepackage{graphicx}
\usepackage{bm}
\usepackage{hyperref}
\def\naa#1{{\bf\nabla}_{#1}^2}
\def\na {{\bf\nabla}}

\def\he#1{$\rm^{#1}He $}
\def\rvec{{\bf r}}

\def\hm#1{{\hbar^2\over #1m}}
\def\hmI#1{{\hbar^2\over #1m_I}}
\def\bra#1{\left\langle #1\right|}
\def\ket#1{\left| #1\right\rangle}
\def\half {\frac{1}{2}}

\def\gddr{\Gamma_{\rm dd}(r)}
\def\xddk{\tilde X_{\rm dd}(k)}
\def\xdek{\tilde X_{\rm de}(k)}
\def\xeek{\tilde X_{\rm ee}(k)}
\def\gddk{\tilde\Gamma_{\rm dd}(k)}
\def\xcck{\tilde X_{\rm cc}(k)}
\def\nddr{N_{\rm dd}(r)}

\def\ndek{\tilde N_{\rm de}(k)}
\def\neek{\tilde N_{\rm ee}(k)}
\def\ncck{\tilde N_{\rm cc}(k)}

\begin{document}

\title{Many--body aspects of positron annihilation in the electron gas}
\author{V. Apaja, S. Denk, and E. Krotscheck}

\affiliation{Institut f\"ur Theoretische Physik, Johannes
Kepler Universit\"at, A 4040 Linz, Austria}

\begin{abstract}
We investigate positron annihilation in the electron gas as a case
study for many--body theory, in particular the optimized Fermi
Hypernetted Chain (FHNC--EL) method. We examine several approximation
schemes and show that one has to go up to the most sophisticated
implementation of the theory available at the moment in order to get
annihilation rates that agree reasonably well with experimental
data. Even though there is basically just one number we look at, the
electron--positron pair distribution function at zero distance, it is
exactly this number that dictates how the full pair distribution
behaves: In most cases, it falls off monotonously towards unity as the
distance increases. Cases where the electron--positron pair
distribution exhibits a dip are precursors to the formation of bound
electron--positron pairs. The formation of electron--positron pairs is
indicated by a divergence of the FHNC--EL equations; from this we can
estimate the density regime where positrons must be localized. This
occurs in our calculations in the range $9.4\le r_s\le 10$, where
$r_s$ is the dimensionless density parameter of the electron liquid.
\end{abstract}
\pacs{78.70.Bj,71.10.Ca} 

\maketitle


\section{Introduction}
\label{sec:Introduction}

The process of electron--positron annihilation has been studied
intensively for several decades. In recent years positron annihilation
spectroscopy has been routinely used for studying the electronic
structures of solids. As far as the two--body process is concerned,
the appropriate theoretical framework is quantum--electrodynamics
(QED).  Differential cross sections and annihilation rates have been
examined for two--particle systems like positronium in much detail,
and can be found in standard textbooks \cite{ItZuber,JauchRohr}.
Coincidence measurements of gamma emission give the angular
correlation of annihilation radiation (ACAR), that yields information
about the electron momenta. As the recent discovery of the
electronically stable bound positron-Li state showed
\cite{rydzhikh-mitroy-97}, positrons in contact with neutral atoms can
also be an interesting few--body system. Since then the list of atoms
binding positrons has become long; the underlying calculations are
usually performed using either the stochastic variational method or
the configuration integration method.

Annihilation of positrons in matter adds a many--body aspect to the
problem. The experimental annihilation rates are by now well
established, we shall compare our results with the data measured by
Weisberg and Berko \cite{expWeisberg} for alkali metals.  In his
pioneering work, Ferrell \cite{Ferrell} gives intuitive formulas for
the annihilation rates. The simplest ``many--body'' methods use single
particle wave functions; in this case, the only true many--body effect
is the Pauli exclusion principle acting between electrons, but {\em
many--body correlations\/} between the interacting particles are {\em
not\/} taken into account. Qualitatively, correlations can be
introduced by applying enhancement factors as described by Brandt
\cite{BrandtVarenna}. A popular method for electron structure
calculations and potentially also for positron systems
\cite{barbiellini-etal-96} is density functional theory (DFT) . In
particular, there are numerous applications of DFT to defects in
solids. In DFT one can write the annihilation rate in terms of the
electron and positron densities, and an enhancement factor
\cite{Kahana63} to account for the excess electron density near the
positron, in other words, to describe electron--positron
correlations.

The enhancement factor has been the subject of many recent studies
\cite{anniArponen,sormann-96,mitroy-barbiellini-02}. Once the
enhancement factor is known, one can apply standard DFT under various
approximate schemes, like the local density approximation (LDA),
generalized gradient approximation (GGA) or the weighted density
approximation (WDA). One can then also evaluate the partial
annihilation rates due to valence and core electrons. The density
functionals are derivatives of known properties of electron gas or
electron--positron mixtures, and their quality has been tested only in
the case of a positron--neutral atom bound system
\cite{mitroy-barbiellini-02}. So far there have not been many attempts
to formulate a microscopic many--body theory that deals with an {\em
inhomogeneous\/} electron gas. A first move to this direction was made
by Stachowiak and Boro{\'n}ski \cite{stachowiak-boronski-01}, who
studied the case of a spherical inhomogeneity in jellium.

Most of the many--body aspects of positron annihilation rates are
reflected in a single number, namely the value of the
electron--positron distribution function, $g^{\rm IB}(r)$, at the
origin. (As a convention, we shall label all two--body quantities that
involve one positron (``impurity'') and one electron (``background'')
with a superscript {\rm IB}.) The annihilation rate of a positron in
{\em homogeneous\/} electron gas can be written in the form
\cite{Ferrell57},
\begin{equation}
{1\over\tau} = {12\over r_s^3}\, g^{\rm IB}(0)\times 10^9 {\rm sec}^{-1}\ ,
\label{eq:tau}
\end{equation}
where $r_s$ is the familiar dimensionless density parameter of the
electron gas. The $r_s$--factor in formula (\ref{eq:tau}) is merely a
geometric factor that takes into account the decreasing probability of
finding an electron at the locations of the positron due to decreasing
electron density. The ``enhancement factor'' $g^{\rm IB}(0)$ accounts
for electron--positron correlations. These can be strong in a metal,
hence $g^{\rm IB}(0)$ can be large. The positron impurity is
delocalized at low $r_s$ and cannot give rise to any appreciable local
enhancement of the electron density. Instead, there is an increased
probability of finding an electron near the positron: This tendency is
visible only in the pair correlations, not in the density. This
explains why annihilation rates computed using a {\em homogeneous\/}
electron gas agree well with the experimental data up to $r_s \sim
5$. On the other hand, the electron density enhancement around a
localized positron can be included, for example, in the spirit of LDA
in DFT, where the electron--positron distribution function computed
for the homogeneous case is multiplied with the spatially varying
electron density \cite{BAL79,BorNi86,puska-nieminen-94}.

The calculation of $g^{\rm IB}(r)$ is a matter of many--body physics,
and the problem of determining $g^{\rm IB}(0)$ is evidently an issue
of {\em short--ranged\/} correlations. Quite appropriately, it was dealt
with within perturbation theory by solving the electron--positron
Bethe--Goldstone equation \cite{Kahana63}. Brown, Jackson, and Lowy
\cite{LoB75,Lowy75} have also examined short--ranged
electron--electron correlations in a Bethe--Goldstone theory and
pointed out the possibility of electron--positron pair formation at
low electron densities (large $r_s$). The theory of positron
annihilation has been developed further by Boro{\'n}ski, Szotek and
Stachowiak \cite{BorSzoSta81} and Rubaszek and Stachowiak
\cite{RubSta84,RubSta88}, and it appears to be able to reproduce the
observed annihilation rates in simple metals.

With the advent of highly re--summed variational techniques, a new
generation of calculations containing vastly richer diagrammatic
structures than Bethe--Goldstone calculations was possible.  As a
physically relevant paradigm for a fermion mixture and electron--hole
liquids, positronic impurities and electron--positron mixtures have
been studied quite extensively. Kallio, Pietil\"ainen and Lantto
\cite{Kallio-Pietilainen-Lantto-82,PitKal83} used the so--called
``quasi--boson'' approximation for the wave function of the electronic
background, which maps the formalism to an effective boson
theory. Closest to our approach are the calculations by Lantto
\cite{Lantto87} and Saarela \cite{MikkoMBVIII}. Compared with the
former one, the present work has an improved diagrammatic summation
and in particular a more consistent treatment of the antisymmetry of
the wave function. Lantto \cite{Lantto87} also employs a simplified
version of the Euler equation which corresponds to our FHNC//0
approximation to be discussed in section \ref{ssec:FHNCcc0}, but the
the energy and the structure functions are evaluated using the full
FHNC equations and thus violates the identities (\ref{eq:Xijlong}),
(\ref{eq:VFermiLong}) and (\ref{eq:XIBdelong}).  Saarela
\cite{MikkoMBVIII} modified the Euler--Lagrange equation of the
pair--distribution function of a charged {\em Bose\/} system by adding
an {\it ad hoc\/} electron--electron potential, such that the equation
reproduces the exact free--fermion distribution in the limit of
infinite density. Although oversimplified and phenomenological, this
approach gives $g^{\rm IB}(0)$ close to the values obtained by
Stachowiak and Lach~\cite{StaL93} and Boro{\'n}ski and
Nieminen~\cite{BorNi86}. Stachowiak {\em et al.} have used HNC theory
in combination with a Hartree--Fock--type approximation and a
self--consistent perturbation of a Jastrow state
\cite{GOS85,STA90,STA97,BorSt98,stachowiak-boronski-banach-00}.

The present work should be considered as an exercise in basic
microscopic many--body techniques. We shall first outline the most
complete version of the optimized Fermi--hypernetted--chain theory
(FHNC--EL) \cite{Mistig}. We will pay special attention to the
full functional optimization of correlation functions, which removes
all ambiguity from the optimization process. The technical details of
our theory for a one--component electron system are described in
Ref.~\onlinecite{annals}, a more recent application to \he3 which
includes the optimization of triplet correlations and the calculation
of proper elementary diagrams may be found in
Ref.~\onlinecite{polish}. We have also recently examined a completely
analogous problem in helium liquids, namely the calculation of
properties of \he4 impurities in \he3 \cite{impu4in3}. We shall then
discuss the impurity theory and derive the relevant Euler
equations. In the following sections, we will lead the reader through
a sequence of plausible approximations in order to determine what it
takes to get the physics right. We will show that the very simple
approximation of a mixture of charged bosons gives a reasonably good
agreement with experimental data. However, a bosonic theory is
unsatisfactory {\em per se,} but it turns out that the first fermionic
corrections make things worse and the agreement is lost. Finally, we
will show that only the full fermion theory produces results which
agree again with experimental data.


\section{Optimized Fermi Hypernetted Chain Method}
\label{sec:FHNC-EL}
\subsection{Variational Wave Function}
\label{ssec:PsiVar}

This section gives a brief survey of the variational theory of a bulk
Fermi liquid; the reader is referred to
Refs. \onlinecite{annals,polish} for details of the theory, and the
diagrammatic definition of all technical quantities.  Since no
confusion can arise for the time being, we will in this section not
spell out the particle species.

The Jastrow--Feenberg theory \cite{FeenbergBook} for a Fermi liquid
assumes a trial wave function of the form
\begin{eqnarray}
        \Psi_{0}(1,\ldots,N) &=& F(\rvec_1,\ldots,\rvec_N)\Phi_0(1,\ldots,N),\\
        F(\rvec_1,\ldots,\rvec_N) &=&
        \exp{1\over2}\left[\sum_{1\le i<j\le N} u_2({\bf r}_i,{\bf r}_j)
        + \sum_{1\le i<j<k\le N}u_3({\bf r}_i,{\bf r}_j,{\bf r}_k)
        + \ldots\right] \,.
\label{eq:JastrowWaveFunction}
\end{eqnarray}
$\Phi_0(1,\ldots,N)$ is a model wave function, normally a
Slater--determinant of plane waves. 
The {\it correlation functions\/} $u_n({\bf r}_1,\ldots,{\bf r}_n)$
are made unique by imposing the {\it cluster--property\/}
\begin{equation}
u_n({\bf r}_1,\ldots,{\bf r}_n)\rightarrow 0
\quad{\rm as}~~|\rvec_i-\rvec_j|\rightarrow\infty\,.
\label{eq:ClusterProperty}
\end{equation}
The wave function (\ref{eq:JastrowWaveFunction}) is not exact; one way
to see this is by realizing that the nodes of the wave function
(\ref{eq:JastrowWaveFunction}) are identical to those of the model
function $\Phi_0(1,\ldots,N)$. In the parlance of Monte Carlo
simulations this would be called a fixed node
approximation. 

\subsection{Fermion HNC equations}
\label{ssec:FHNC}
Two components are essential for the execution of the
Jastrow--Feenberg variational theory: The first is the development of
cluster--expansion and resummation methods for the pair distribution
function in the homogeneous case,
\begin{equation}
        g(r) =
        {N(N-1)\over \rho^2}
        {\int d^3r_3\ldots d^3r_N\left|\Psi_0(1,\ldots,N)\right|^2
        \over
        \int d^3r_1\ldots d^3r_N\left|\Psi_0(1,\ldots,N)\right|^2}\,,
\label{eq:rhotwo}
\end{equation}
where $r=r_{12} = \left|{\bf r}_1-{\bf r}_2\right|$; spin--summations
are tacitly implied. The second component of the theory is the {\it
optimization\/} of the correlation functions by minimization of the
total energy
\begin{equation}
{\delta\over \delta\;u_n({\bf r}_1,\ldots,{\bf r}_n)}
{\left\langle\Psi_0\right| H \left |\Psi_0\right\rangle
\over
{\left\langle\Psi_0\mid\Psi_0\right\rangle}} = 0
\end{equation}
for the Hamiltonian
\begin{equation}
H       = -\sum_{i=1}^N{\hbar^2\over 2m}\nabla_i^2
        + \sum_{1\le i< j\le N} v(|{\bf r}_i-{\bf r}_j|)
\label{Hamiltonian}
\end{equation}
where, in our case $v(r) = e^2/r$ is the Coulomb interaction.

Let us first turn to the pair distribution function $g(r)$,
specifically to the FHNC equations determining $g(r)$ from a given
pair correlation function $u_2(r)$. Pair correlations are the most
important ones, and in the case of electrons one usually neglects
triplet correlations altogether. Justification for this stems from the
study of triplet correlations in the 2D and 3D charged Bose
gas \cite{bosegas} and verified by Monte Carlo calculations
\cite{3DBackflow}, note that triplet correlations are {\it not\/} the
same as propagator corrections \cite{polish} which have occasionally
been confused with Feynman--Cohen backflow.

The FHNC equations are a set of four configuration--space, and four
momentum--space equations formulated in terms of ``nodal'' $N_{\rm
ij}(r)$ and ``non--nodal'' diagrams $X_{\rm ij}(r)$ that are, in turn,
characterized by their exchange structure $\{\rm ij\}\in\{\rm
dd,de,ee,cc\}$.  Input to the equations is the pair correlation
function $u_2(r)$, the Slater exchange function $\ell(x) = 3(\sin\,x -
x\cos\,x)/x^3$, and a set of ``elementary diagrams'' $E_{\rm ij}(r)$
\cite{Johnreview} that must be calculated one by one.

The coordinate--space equations are
\begin{eqnarray}
\gddr           &=& X_{\rm dd}(r) + N_{\rm dd}(r)
\nonumber\\
                &=& \exp\left[u_2(r)+N_{\rm dd}(r) + E_{\rm dd}(r)\right]
                -1 \,,
\label{eq:Gddr}\\
X_{\rm de}(r)   &=& \left[1+ \gddr\right]
                \left[N_{\rm de}(r) +E_{\rm de} (r)\right]-N_{\rm de}(r)\,,
\nonumber\\
X_{\rm ee}(r)   &=& \left[1+ \gddr\right]
                \left[- {1\over\nu} L^2 (r) + N_{\rm ee} (r)
                + E_{\rm ee} (r)\right]
                -N_{\rm ee}(r)  \nonumber\\
                &+&\left[1+ \gddr)\right]
                \left[N_{\rm de} (r) + E_{\rm de} (r)\right]^2\,,
\nonumber\\
X_{\rm cc}(r)   &=& -\frac 1 \nu \gddr L(r) + E_{\rm cc}(r)
\nonumber
\end{eqnarray}
where $\nu$ is the degree of degeneracy of the single--particle
states, $L(r) = \ell(k_Fr)-\nu\left[N_{\rm cc}(r) + E_{\rm
cc}(r)\right]$, and $k_F$ is the Fermi wave number.  The ``nodal''
quantities $N_{\rm ij}(r)$ are constructed in momentum space according
to
\begin{eqnarray}
\tilde N_{\rm dd}(k)    &=& {\tilde X_{\rm dd}(k)
        \over \left[1 - \tilde X_{\rm de}(k)\right]^2 - 
        \left[ 1 + \tilde X_{\rm ee}(k)\right]\tilde X_{\rm dd}(k)}
        - \tilde X_{\rm dd}(k)\,,
\nonumber\\
\ndek   &=& {1- \xdek -\xddk \over \left[1 - \xdek\right]^2 - 
        \left[ 1 + \xeek\right]\xddk}
        - 1 -\xdek\,,\nonumber\\
\neek   &=& { \xddk + 2\xdek +\xeek -1
                \over [1 - \xdek]^2 -  [ 1 + \xeek ]\xddk }+1 -\xeek\,,
\label{eq:NodalF}\\
\ncck   &=&
        - \xcck \left[{\tilde l (k) /\nu - \xcck \over 1 - \xcck }\right]\,.
\nonumber
\end{eqnarray}
We have above used the convention of defining a dimensionless
Fourier--transform as $\tilde f(k) \equiv \left[f(r)\right]^{\cal
F}(k) \equiv \rho\int d^3 r f(r) e^{i {\bf
k}\cdot{\bf r}}$.

The pair distribution function can then be constructed from the
above quantities:
\begin{eqnarray}
        g(r)
        &=&\left[1+ \gddr\right]
        \left\{ - {1\over\nu} L^2 (r) + N_{\rm ee}(r) + E_{\rm ee} (r)
 +\left[1 + N_{\rm de}(r) + E_{\rm de} (r)\right]^2 \right\}\,.
\label{eq:gFHNC}
\end{eqnarray}
The static structure function is equally important than the pair 
distribution function and has the relatively simple form
\begin{eqnarray}
        S(k) &=& 1 + \left[g(r)-1\right]^{\cal F}(k)\nonumber\\
        &=& { 1 + \xeek \over [1 - \xdek]^2 -  [ 1 + \xeek ]\xddk }\,.
\label{eq:SFHNC}
\end{eqnarray}
For further reference, we also introduce the quantity
\begin{equation}
S_{\rm d}(k) =  { 1 -\xdek \over [1 - \xdek]^2 -  [ 1 + \xeek ]\xddk }\,,
\label{eq:Sdir}
\end{equation}
and note the relationship
\begin{equation}
\gddk
=  {\xddk \over [1 - \xdek]^2 -  [ 1 + \xeek ]\xddk }\,.
\label{eq:Gammadd}
\end{equation}
The representation (\ref{eq:gFHNC}) of $g(r)$ contains an explicit factor
\begin{equation}
1+\gddr
= \exp\left[u_2(r)+N_{\rm dd}(r) + E_{\rm dd}(r)\right]\,.
\end{equation}
This form is therefore the natural choice when working in coordinate
space and focusing on the strong short--range correlation structure.
On the other hand, when we consider the static structure function
$S(k)$, the expression (\ref{eq:SFHNC}) is the more useful one.

The na\"\i ve implementation \cite{Fantoni,Zabolitzky} of the FHNC
equations, referred to as FHNC//0 approximation, would suggest, in
analogy to the boson theory, to start with the omission of the
``elementary diagrams'' $E_{\rm ij}(r)$ and include these, order by
order, as {\it quantitative\/} improvements as the theory is moved to
the next level. In the FHNC//n approximation one keeps ``elementary''
diagrams up to the $n$--point diagram. However \cite{Mistig}, such a
procedure violates the exact features
\begin{eqnarray}
        \xdek =& {\cal O}(k) \qquad&{\rm as}\quad k\rightarrow
        0^+\,,
\nonumber\\
        1 + \tilde X_{\rm ee}(k) =& S_{\rm F}(k) + {\cal O}(k^2)\qquad
        &{\rm as}\quad k\rightarrow 0^+\,.
\label{eq:Xijlong}
\end{eqnarray}
These properties originate from the Pauli principle and are
particularly important for the optimization problem \cite{EKVar}.
They imply the cancellation of ``elementary'' and ``non--elementary''
exchange diagrams; in other words there exist classes of so--called
``elementary'' exchange diagrams that must not be neglected.  We have
dealt with these diagrams in an approximate way, dubbed as
``//C''--approximation, which has been described and justified in
Ref. \onlinecite{polish}.


\bigskip
\subsection{Background Energy calculation}
\label{sec:FEnergy}

The first step in deriving the Euler equations is the calculation of
the energy functional.  An important manipulation is the use of the
{\it Jackson--Feenberg\/} identity
\begin{equation}F \nabla^2 F =
        \half (\nabla^2 F^2+ F^2\nabla^2) + \half F^2
        \left[\bm\nabla,\left[\bm\nabla ,\ln F\right]\right]
        - {1\over 4} \left[\bm\nabla,\left[\bm\nabla, F^2 \right]\right]\,,
\label{eq:JFIdentity}
\end{equation}
which shows that the expectation value of the kinetic energy can be
divided into three parts,
\begin{equation}
\langle \hat T\rangle
=       T_{\rm F} - {N \hbar^2\rho \over 8m}\int d^3r\; g(r)
        \nabla^2 u_2(r)
        + T_{\rm JF}\label{eq:EkinFermi}\,.
\end{equation}
Here $T_{\rm F}$ is the kinetic energy of the free Fermi gas, and
$T_{\rm JF}$ is a kinetic energy term that is solely due to exchanges.
We can write this term as
\begin{equation}
T_{\rm JF}      \equiv {\hbar^2\over 8m}
        \sum_i{\bra{\Phi_0} \left[ \bm\nabla_i,\left[\bm\nabla_i, F^2\right]
        \right] \ket{\Phi_0}
                \over \bra{\Phi_0} F^2 \ket{\Phi_0}}
        \equiv {\hbar^2 N\over 8m }\int d^3r\,\nabla_\ell^2\rho_1(\rvec)\,.
\label{eq:TJF}
\end{equation}
With $\bm\nabla_\ell$ we mean (indicated by the subscript $\ell$) a
gradient operator that acts on the Slater determinant
only. Operationally, Eq. (\ref{eq:TJF}) is to be understood as
follows: First one calculates --- disregarding the fact that the
density is uniform --- a cluster expansion of the one--body density
$\rho_1(\rvec)$. The operator $\nabla_\ell^2$ then differentiates
{\it the exchange lines only\/} that are attached, in such a cluster
expansion, to the external point. This replaces the incoming and
outgoing exchange lines $\ell(\left|{\bf r}-{\bf r}_i\right|k_F)
\ell(\left|{\bf r}-{\bf r}_j\right|k_F)$ at the reference point by
$\left({\hbar^2/8m}\right)\nabla_{\bf r}^2 \ell(\left|{\bf r}-{\bf
r}_i\right|k_F) \ell(\left|{\bf r}-{\bf r}_j\right|k_F)$.  Finally,
one takes the limit of the uniform system, and integrates over the
configuration space of the last particle.

Combining Eqs. (\ref{eq:EkinFermi}) and (\ref{eq:TJF}) with the
potential energy provides us with the starting point for further
manipulations,
\begin{equation}
{E\over N} = {T_F\over N} + {\rho\over 2}\int d^3 r\; g(r) v_{\rm JF}(r) +
{T_{\rm JF}\over N}\,,
\label{eq:EJF}
\end{equation}
where
\begin{equation}
v_{\rm JF}(r) = v(r) - {\hbar^2\over 4m} \nabla^2 u_2(r)
\label{eq:vJF}
\end{equation}
is the {\it Jackson--Feenberg effective interaction.\/}

\subsection{Fermion Euler equations}
\label{ssec:FermiEuler}

The formal manipulations to derive an Euler equation for the optimal
pair correlations are almost identical to the ones carried out for
bosons. The variation with respect to the pair correlation function
consists of two terms: One comes from the variation with respect to
the pair correlation function $u_2(r)$ appearing in the
Jackson--Feenberg interaction $v_{\rm JF}(r)$, and the second one is
due to the variation of the pair distribution function with respect to
$u_2(r)$ {\it and\/}  the variation of $T_{\rm JF}$:
\begin{equation}
        {\hbar^2\over 4m}\nabla^2 g(r) =
        \int d^3r'\, v_{\rm JF}(r')
        {\delta g(r')\over \delta u_2(r)}
        + {2\over\rho}{\delta \over \delta u_2(r)}{T_{\rm JF}\over N}
\equiv g'(r)\,.
\label{eq:EulerPrimF}
\end{equation}
The contribution from the first term on the right hand side of
Eq. (\ref{eq:EulerPrimF}) to $g'(r)$ is calculated in complete analogy
to the Bose case by replacing, in turn, each correlation line
$\exp\,\left[u_2(r_{ij})\right]-1$ by
$\exp\,\left[u_2(r_{ij})\right]\,v_{\rm JF}(r_{ij})$. The second term
is calculated recalling the graphical construction scheme of $T_{\rm
JF}$ described above, and applying the same procedure to a graphical
expansion of $g(r)$. Thus, the contribution to $g'(r)$ originating
from $T_{\rm JF}$ is obtained by replacing in $g(r)$, in turn, every
connected pair of exchange lines $\ell(r_{ij}k_F) \ell(r_{ik}k_F)$ by
$({\hbar^2/8m})\nabla_i^2\ell(r_{ij}k_F) \ell(r_{ik}k_F)$.  Following
this construction scheme, one derives a set of eight linear equations,
the FHNC'--equations, corresponding to the eight FHNC equations
(\ref{eq:Gddr}), (\ref{eq:NodalF}) \cite{annals}, in which the
Jackson--Feenberg effective potential and the differentiated exchange
functions act as driving terms.

To derive a form of the fermion Euler equations that is useful for a
numerical implementation, we write the Euler equation
(\ref{eq:EulerPrimF}) in momentum space as
\begin{equation}
        {1\over2}t(k)\left[S(k)-1\right] + S'(k) = 0\,,
\label{eq:FermiEulerK}
\end{equation}
where $t(k) \equiv \hbar^2k^2/2m$.  The $S'(k)$ is a linear
combination of the non--nodal quantities $\tilde X'_{ij}(k)$
\begin{equation}
        S'(k) = \sum_{ij\in \{\rm dd,de,ee\}}
                {\partial S (k)\over \partial X_{\rm ij} (k)}
                \tilde X'_{\rm ij} (k)
\label{eq:SprimeF}
\end{equation}
\noindent
where the $\tilde X'_{\rm ij} (k)$ are constructed from the non--nodal
quantities $\tilde X_{\rm ij} (k)$ analogously to the construction of
$g'(r)$ from $g(r)$ described above. Next, we define three effective
interactions in the $\rm dd$--, $\rm de$-- and $\rm ee$--channels as
\begin{eqnarray}
\tilde V_{\rm dd}(k) & =& \tilde X'_{\rm dd}(k) - {1\over2}\, t(k)\xddk\,,
\nonumber\\
\tilde V_{\rm de}(k) & =& \tilde X'_{\rm de}(k)\,,\label{eq:VphFermi}\\
\tilde V_{\rm ee}(k) & =& \tilde X'_{\rm ee}(k) + {1\over 2}\, t(k)
\xeek\,.
\nonumber
\end{eqnarray}
The quantity $\tilde V_{\rm dd}(k)$, defined in
Eq. (\ref{eq:VphFermi}), may be identified with the ``direct
interaction'' of the Babu--Brown theory \cite{BabuBrown} of the
quasiparticle interaction.  From Eqs. (\ref{eq:Xijlong}), the
effective interactions inherit the long--wavelength properties
\begin{eqnarray}
        \tilde V_{\rm de}(k) &= {\cal O}(k)
        &\qquad{\rm as}\qquad k \rightarrow 0^+\,,
        \label{eq:VFermiLong}\\
        \tilde V_{\rm ee}(k) & = {\cal O}(k^2)
        &\qquad{\rm as}\qquad k \rightarrow 0^+\,.\nonumber
\end{eqnarray}

Using the representation (\ref{eq:SFHNC}) for the calculation of
$S'(k)$ via Eq. (\ref{eq:SprimeF}) and eliminating the $X'_{\rm
ij}(k)$ in favor of the $V_{\rm ij}(k)$ lets us rewrite
\begin{equation}
S'(k) = S^2 (k) \tilde V_{\rm dd}(k) +
2S(k)S_{\rm d} (k) \tilde V_{\rm de} (k)+S_{\rm d}^2 (k)
\tilde V_{\rm ee} (k) + 
{1\over2}\,t(k)\left[S_{\rm d}^2(k)-S(k)\right]\,.
\label{eq:SprimeFermi}
\end{equation}
Inserting this expression for $S'(k)$ in the Euler equation
(\ref{eq:FermiEulerK}) lets us express $S(k)$ in terms of the
three effective interactions $\tilde V_{\rm ij}(k)$.

The second step in the derivation of the Euler equation is to eliminate
the pair correlation function $u_2(r)$ by using the FHNC equation
(\ref{eq:Gddr})
\begin{eqnarray}
        -\hm4\left[1+\gddr\right]\nabla^2 u_2(r) =& -&\hm4\nabla^2\gddr +
        {\hbar^2\over m}\left|\bm\nabla\sqrt{1+\gddr}\right|^2\nonumber\\
        \qquad\qquad&+&\hm4\left[1+\gddr\right]\left[\nabla^2\nddr+
        \nabla^2 E_{\rm dd}(r)\right]\,.
\label{eq:Del2u}
\end{eqnarray}
Using Eq. (\ref{eq:Del2u}), one can rewrite the effective interactions
$V_{\rm ij}(r)$ in coordinate space entirely in terms of the
distribution functions, the yet unspecified sets of elementary
diagrams $E_{\rm ij}(r)$ and their ``primed'' counterparts $E'_{\rm
ij}(r)$.  The resulting equations are lengthy and not very
illuminating; they have been spelled out in Refs. \onlinecite{annals}
and \onlinecite{polish}.  For the purpose of comparison with the
impurity results, we display the coordinate space form of the direct
interaction
\begin{eqnarray}
V_{\rm dd}(r)&=& \left[1+ \gddr\right]
\left[v(r) +\hm4\nabla^2E_{\rm dd}(r) + E'_{\rm dd}(r)\right]
\nonumber\\
&+& {\hbar^2\over m}\left|\bm\nabla\sqrt{1+\gddr}\right|^2 +
\gddr w_{\rm I}(r)\,,\\\label{eq:VddFermi}
w_{\rm I}(r) &=&
\hm4\nabla^2\nddr+N_{\rm dd}'(r)\,.
\label{eq:windFermi}
\end{eqnarray}
To calculate the effective interactions $\tilde V_{\rm ij}(k)$ ($(\rm{
ij})\in \{{\rm dd, de, ee, cc}\})$ (or $\tilde X'_{\rm ij}(k)$) and
the fermion analog of the induced potential $w_{\rm I}(r)$, one
must also calculate the ``primed'' analogs of the ``nodal'' diagrams
$N_{\rm ij}(r)$. These quantities may be found in
Refs. \onlinecite{annals} and \onlinecite{polish}, they are needed for
the numerical optimization, but we will not need them in the further
discussion. The ${\rm dd}$--elementary diagrams have no special
features; their omission can cause quantitative changes in the
final answer, but does, unlike the ${\rm de}$ and ${\rm ee}$
``elementary'' diagrams, not change the analytic structure and
the properties of the solutions. 

\section{Impurity correlations}
\label{sec:FHNC-ELI}

In this section, and further on, we also must spell out the particle
species, which may be an electron (``background particle'', referred to by
a superscript ``B'' ) or a positron (``impurity particle'', referred to by
a superscript ``I''). It is not necessary to label those quantities
that were introduced in the last section and that refer only to
background particles. We will keep the formulation general in the
sense that the impurity mass $m_I$ is arbitrary, as well as the
interaction between an ``impurity'' (positron) and a ``background'' (electron)
particle. The Hamiltonian for the full system including the impurity is
given by

\begin{equation}
        H^I = -\hmI2\naa0 
        - \sum_{j=1}^N\hm2\naa j
        + \sum_{j=1}^N v^{\rm IB}(|\rvec_0 - \rvec_j|)
        + \mathop{\sum_{j,k=1}}_{j<k}^N
        v(|\rvec_j - \rvec_k|)\,.
\label{Himp}
\end{equation}
As a convention, the impurity particle coordinate is $\rvec_0$. In the
present case of a positron in an electron gas, the
impurity--background interaction $v^{\rm IB}(r)$ only differs in the
overall sign from the background--background interaction $v(r)$, which
is simply the repulsive Coulomb potential. The impurity mass $m_I$ is
equal to the background mass $m$, but for a better insight into the
problem it is helpful to keep $m_I$.

The formulation of the FHNC--EL equations for a single impurity
follows essentially the same path as the formulation of the background
equations, namely
\begin{itemize}
\item{} Define a variational wave function
\begin{equation}
\Psi_0^{\rm IB}(0,1,\ldots,N) =
\exp\left[{1\over2}\sum_{i=1}^N u_2^{\rm IB}(\rvec_0,\rvec_i)\right]
\Psi_0(1,\ldots,N)\,,
\label{eq:ImpurityWaveFunction}
\end{equation}
\item{} Derive a set of FHNC equations,
\item{} Derive the corresponding Euler equation using the ``prime equation''
technique, and
\item{} Reformulate the Euler equation in terms of distribution
functions, thereby eliminating any reference to the correlation
function $u_2^{\rm IB}(r_{ij})$.
\end{itemize}
Compared to finite--concentration mixtures, the derivation is
simplified because there are no exchanges connected to impurity
coordinates.

\subsection{FHNC equations for one impurity}
\label{ssec:FHNC_I}

The (F)HNC technique is well established in earlier work, so there is
no need to go through the details of the derivations here. The aspect
that distinguishes fermions from bosons are the combinatorial rules
and long--wavelength properties discussed above. In a mixture,
exchanges can only take place between particles of the same species;
this implies in the dilute limit that exchanges occur only between
``background'' particles. Moreover, in the dilute limit of a mixture
{\it only one\/} impurity can occur in each diagrammatic quantity.

Since the impurity cannot be involved in any exchange, we have only
two FHNC equations: The equations describing the parallel connections
between external coordinates are
\begin{eqnarray}
\Gamma^{\rm IB}_{\rm dd}(r) &=& X^{\rm IB}_{\rm dd}(r) + N^{\rm IB}_{\rm dd}(r)
= \exp{\left(u_2^{\rm IB}(r) + N^{\rm IB}_{\rm dd}(r) +
E^{\rm IB}_{\rm dd}(r)\right)} -1
\label{eq:I_hypernet},\\
X^{\rm IB}_{\rm de}(r) &=& \left[1 + \Gamma^{\rm IB}_{\rm dd}(r) \right]
\,\left[E^{\rm IB}_{\rm de}(r) + N^{\rm IB}_{\rm de}(r)\right]
-N^{\rm IB}_{\rm de}(r)\nonumber
\end{eqnarray}
while the chain connections are best written in momentum space,
\begin{eqnarray}
\tilde N^{\rm IB}_{\rm dd}(k) &=&
\left( S_{\rm d}(k)-1 \right) \,
\tilde X^{\rm IB}_{\rm dd}(k) +
\tilde \Gamma_{\rm dd}(k)\,\tilde X^{\rm IB}_{\rm de}(k)\,,
\nonumber\\
\tilde N^{\rm IB}_{\rm de}(k) &=& \left(S(k) - S_{\rm d}(k)\right)
\,\tilde X^{\rm IB}_{\rm dd}(k) + \left(S_{\rm d}(k)
-1 - \tilde \Gamma_{\rm dd}(k)\right)
\,\tilde X^{\rm IB}_{\rm de}(k)\,.\label{eq:I_chain}
\end{eqnarray}
From these quantities, we can construct the impurity--background
distribution function
\begin{equation}
g^{\rm IB}(r) = \left[1 + \Gamma^{\rm IB}_{\rm dd}(r)\right]
\left[1 + N_{\rm de}^{\rm IB}(r) + E_{\rm de}^{\rm IB}(r)\right]\,.
\label{eq:gIB}
\end{equation}
The long--wavelength properties corresponding to the identities
(\ref{eq:Xijlong}) apply only for the background coordinates.  Since
the exchange structures of $X^{\rm IB}_{\rm de}(r)$ and $X_{\rm
de}(r)$ are the same, we have the long--wavelength limit
\begin{equation}
\tilde X^{\rm IB}_{\rm de}(k) = {\cal O}(k) \qquad{\rm as}\quad k\rightarrow
        0^+\,.
\label{eq:XIBdelong}
\end{equation}
To abbreviate the equations, we found it convenient to define the
quantity
\begin{equation} 
\tilde {\cal X}^{\rm IB}(k) \equiv  \tilde X^{\rm IB}_{\rm dd}(k)
+ {S_{\rm d}(k)\over S(k)} \tilde X^{\rm IB}_{\rm de}(k)\,.
\label{eq:calXdef}
\end{equation}
A few relations are useful in the derivation of the resulting
equations:
\begin{eqnarray}
S^{\rm IB}(k) &\equiv& \left[g^{\rm IB}(r)-1\right]^{\cal F}(k)
= S(k)\tilde X^{\rm IB}_{\rm dd}(k) + S_{\rm d}(k)
\,\tilde X^{\rm IB}_{\rm de}(k) = S(k)\tilde {\cal X}^{\rm IB}(k)\,,
\nonumber\\
\tilde \Gamma^{\rm IB}_{\rm dd}(k) &=&
S_{\rm d}(k) \,\tilde X^{\rm IB}_{\rm dd}(k) +
\tilde \Gamma_{\rm dd}(k)\,\tilde X^{\rm IB}_{\rm de}(k)
= S_{\rm d}(k) \,\tilde{\cal X}^{\rm IB}(k) 
- {\tilde X^{\rm IB}_{\rm de}(k)\over 1+\xeek}\,,
\label{eq:GammaddI}\\
\tilde \Gamma_{\rm dd}(k) &=&  {S_{\rm d}^2(k)\over S(k)} 
- {1\over 1+\tilde X_{\rm ee}(k)}\,.\nonumber
\end{eqnarray}

\subsection{Euler equations for one impurity}

For the determination of existence and stability of the solutions of
the optimization problem, it is again useful to formulate the Euler
equation in momentum space; the inclusion of the appropriate classes
of ``elementary'' exchange diagrams always guarantees the proper
short--distance behavior. Useful abbreviations are $t_I(k) = {\hbar^2
k^2/ 2m_I}$ and the Feynman spectrum of the background,
$\hbar\omega(k) = {t(k)/S(k)}$. The formal Euler equation for the
impurity--background correlations is \cite{MixMonster}, in analogy to
Eq. (\ref{eq:FermiEulerK}).

\begin{equation}
{1\over4} (t_I(k)+t(k)) S^{\rm IB}(k) + {S'}^{\rm IB}(k) = 0\,.
\label{eq:FermiEuler_I}
\end{equation}
The remaining manipulations are to carry out the ``priming'' operation
on the 
impurity FHNC equations and to formulate the
equations in a reasonably plausible form.  This can be done in many
ways, and the ultimate choice of the formulation depends to some
extent on the iteration path adopted for the numerical solution.
Formally, we can define --- in analogy to $\tilde {\cal X}^{\rm
IB}(k)$ introduced above --- and to Eqs. (\ref{eq:VphFermi})
\begin{equation} 
\tilde {\cal V}^{\rm IB}(k) \equiv \tilde {\cal X}^{\rm 'IB}(k)
- {1\over4}\, (t(k) + t_I(k))\tilde {\cal X}^{\rm IB}(k)
\label{eq:calVdef}
\end{equation}
and rewrite the impurity Euler equation (\ref{eq:FermiEuler_I})
as
\begin{equation}
\tilde {\cal X}^{\rm IB}(k)
= - 2 {\tilde{\cal V}^{\rm IB}(k) \over t_I(k) + \hbar\omega(k)}\,.
\label{eq:eulerI}
\end{equation}
This representation of the Euler equation is formally identical to the
Euler equation for impurities in Bose liquids. Of course, we still
need to derive working formulas for calculating the quantity
$\tilde{\cal V}^{\rm IB}(k)$.

\subsection{Induced Interactions}

From the definitions (\ref{eq:calVdef}) and
(\ref{eq:calXdef}), we can write formally
\begin{eqnarray}
\tilde{\cal V}^{\rm IB}(k) &=& \tilde V_{\rm dd}^{\rm IB}(k)
+  {S_{\rm d}(k)\over S(k)} \tilde V^{\rm IB}_{\rm de}(k)
+ 
\left({S_{\rm d}(k)\over S(k)}\right)'\tilde X^{\rm IB}_{\rm de}(k)\nonumber\\
 &=& \tilde V_{\rm dd}^{\rm IB}(k)
+  {S_{\rm d}(k)\over S(k)} \tilde V^{\rm IB}_{\rm de}(k)
- {\tilde X^{\rm IB}_{\rm de}(k)\over 1+\xeek}
\left[\tilde V_{\rm de}(k)
+  {S_{\rm d}(k)\over S(k)} \tilde V_{\rm ee}(k)\right]\nonumber\\
&+& {t(k)\over 2}{S_{\rm d}(k)\over S(k)}{\tilde X^{\rm IB}_{\rm de}(k)
\xeek\over 1+\xeek} 
\end{eqnarray}
with (note that we deviate, for convenience and consistency
with the definition (\ref{eq:calVdef}) 
slightly from the definitions (\ref{eq:VphFermi}))
\begin{eqnarray}
\tilde V_{\rm dd}^{\rm IB}(k) & =& \tilde X^{'{\rm IB}}_{\rm dd}(k)
- {1\over 4}\left(t_I(k) + t(k)\right)\tilde X^{\rm IB}_{\rm dd}(k)\,,
\nonumber\\
\tilde V_{\rm de}^{\rm IB}(k) & =& \tilde X^{'{\rm IB}}_{\rm de}(k)
- {1\over 4}\left(t_I(k) + t(k)\right)\tilde X^{\rm IB}_{\rm de}(k)\,.
\end{eqnarray}

\noindent The calculation of $ V_{\rm dd}^{\rm IB}(r)$ is identical to the one
for bosons and mixtures:
\begin{eqnarray}
V_{\rm dd}^{\rm IB}(r) &=& \left[1+\Gamma^{\rm IB}_{\rm dd}(r)\right]
\left[v^{\rm IB}(r) + \Delta V_e^{\rm IB}(r)\right]
\label{eq:VddIBdef}
\nonumber\\
&+& \left[{\hbar^2\over 2m}
+{\hbar^2\over 2m_I}\right]\left|\bm\nabla \sqrt{1+\Gamma^{\rm IB}_{\rm
dd}(r)}\right|^2  + \Gamma^{\rm IB}_{\rm dd}(r) w_I^{\rm IB}(r)
\end{eqnarray}
with the induced interaction
\begin{equation}
\tilde w_I^{\rm IB}(k) = \tilde N^{'{\rm IB}}_{\rm dd}(k) -
{1\over4}\left(t_I(k)+t(k)\right) \tilde N^{\rm IB}_{\rm dd}(k)
\end{equation}
and the elementary--diagram correction
\begin{equation}
\Delta V_e^{\rm IB}(r) = \left[{\hbar^2\over 8m} +
{\hbar^2\over 8m_I}\right]\nabla^2 E_{\rm dd}^{\rm IB}(r) +
E_{\rm dd}^{'\rm IB}(r)\,.
\end{equation}
Finally, we need
\begin{eqnarray}
V^{\rm IB}_{\rm de}(r) &=& E^{'{\rm IB}}_{\rm de}(r)\,\left( 1 +
 \Gamma^{\rm IB}_{\rm dd}(r) \right) + \Gamma^{'{\rm IB}}_{\rm
 dd}(r)\,\left(E^{\rm IB}_{\rm de}(r) + N^{\rm IB}_{\rm de}(r) \right)
\nonumber\\
 &+& \Gamma^{\rm IB}_{\rm dd}(r)\,N^{'{\rm IB}}_{\rm de}(r)
+\left(\hmI8 + \hm8\right) \naa{} X^{\rm IB}_{\rm de}(r)
\,,
\label{eq:Xdeprime}
\end{eqnarray}
where the ${\rm de}$--elementary diagrams must be chosen to
guarantee the property $\tilde V^{\rm IB}_{\rm de}(k)\rightarrow 0$
as $k\rightarrow 0^+$, {\it cf.\/} Eq. (\ref{eq:XIBdelong}).

For the calculation of the remaining ingredients, it is
most convenient to start with the  effective interaction
of correlated basis functions (CBF) theory
\begin{equation}
\tilde V_{\rm eff}^{\rm IB}(k) = \Gamma^{'{\rm IB}}_{\rm dd}(k)
-\,{1\over4}(t_I(k)+t(k))\tilde \Gamma^{\rm IB}_{\rm dd}(k)\,.
\label{eq:CBFint}
\end{equation}
Again, using background and impurity Euler equations, one can
write this quantity as
\begin{eqnarray}
 \tilde V_{\rm eff}^{\rm IB}(k)
=&-&{1\over2}\left[t_I(k)+{t(k)\over 1+\xeek}\right]
\tilde \Gamma^{\rm IB}_{\rm dd}(k)\nonumber\\
&-&{1\over 1+\xeek}\,\left[\tilde V^{\rm IB}_{\rm de}(k) +
S^{\rm IB}(k)\tilde V_{\rm de}(k)
 + \tilde \Gamma^{\rm IB}_{\rm dd}(k)\tilde V_{\rm
ee}(k)
+{1\over2}(t_I(k)+t(k))\tilde X_{\rm de}^{\rm IB}(k)\right]\,.\nonumber\\
\end{eqnarray}
From this, we can obtain, for example, the induced interaction
\begin{equation}
\tilde w_I^{\rm IB}(k) = \tilde V_{\rm eff}^{\rm IB}(k) -
\tilde V_{\rm dd}^{\rm IB}(k)
\end{equation}
and, of course, $\tilde \Gamma^{'{\rm IB}}_{\rm dd}(k)$
which is needed for the $de$--equation, and  $\tilde N^{'{\rm IB}}_{\rm de}(k)$
which is now easily obtained from
\begin{eqnarray}
\tilde N^{'{\rm IB}}_{\rm de}(k) &=& S^{'\rm IB}(k)
-\tilde \Gamma^{'\rm IB}_{\rm dd}(k) - \tilde X^{'\rm IB}_{\rm
de}(k)\nonumber\\
&=&-{1\over4}(t_I(k)+t(k))\left(S^{\rm IB}(k)+
\tilde \Gamma^{\rm IB}_{\rm dd}(k)+\tilde X^{\rm IB}_{\rm dd}(k)\right) -
\tilde V_{\rm eff}^{\rm IB}(k) - \tilde V^{\rm IB}_{\rm de}(k)\,.
\nonumber
\end{eqnarray}
With this, we have derived a complete set of equations that can be
solved by iteration.

\subsection{Impurity Energetics}
Another physical quantity of interest is the 
chemical potential of the  positron impurity, which is the energy gained
or lost by adding one impurity particle into the liquid, {\it i.e.\/}
the energy difference
\begin{eqnarray}
        \mu^I &=& E_{N+1} - E_N
\nonumber\\
        &=& {\langle\Psi^I\vert H^I \vert \Psi^I\rangle
                \over \langle\Psi^I\vert\Psi^I\rangle}
        -{\langle\Psi\vert H \vert \Psi\rangle\over
        \langle\Psi\vert\Psi\rangle}\,.
\label{eq:chemi}
\end{eqnarray}
In the calculation of the impurity chemical potential from the
definition (\ref{eq:chemi}) we must include, besides the explicit terms
containing impurity distribution functions, also the {\it changes\/}
in the background distribution and correlation functions due to the
presence of an impurity.  These changes are of the order of $1/N$ and
therefore cause a change of order unity in the positron correlation
energy.  For brevity, we suppress here the
contribution from triplet calculations; these corrections are
already discussed in Ref. \onlinecite{impu4in3}. 
The impurity chemical potential is,
at the pair correlation level, given by
\begin{eqnarray}
        \mu^I&&
        =\rho \int d^3r \;\left[[g^{\rm IB}(r)-1]v^{\rm IB}(r)
        -g^{\rm IB}(r)\left(\hmI8 + \hm8\right)\naa{}
        u_2^{\rm IB}(r)\right]
\nonumber\\
&&      +\, {\rho^2\Omega\over2}\Delta\left\{
        \int d^3r\;\left[ (g(r)-1)v(r)
        - {\hbar^2\over4m}g(r)\naa{}
        u_2(r)\right]\right\} + \Delta T_{\rm JF}
\label{eq:muI}
\end{eqnarray}
where the $\Delta$'s in the second line of Eq.
(\ref{eq:muI}) indicate that we take the difference of the expressions
calculated for the full system {\it minus\/} the same expression for
the pure background.

The pair correlations between background and impurity particles are
determined by optimization. Having a relationship between the pair
distribution functions $g^{ij}(r)$ and the pair correlation functions
$u_2^{ij}(r)$ allows us to {\it choose\/} which one of these four
quantities we consider to be the independent ones. The most convenient
choice is to use the ``dressed'' correlation functions $\gddr$ and
$\Gamma_{\rm dd}^{\rm IB}(r)$ since all other diagrammatic quantities
can be defined in terms of these functions, whereas $u_2(r)$ appears
explicitly only in the coordinate space equation (\ref{eq:Gddr}). In
other words, we consider $u_2(r)$ as a functional of $\gddr$, the
impurity density $\rho^I$, and $\rho$. As mentioned earlier, the
consideration is further simplified by the fact that the impurity
cannot be involved in any exchange.

Since we need the chemical potential $\mu^I$ only to leading order in
the impurity density, the optimization conditions for the background
can be used to simplify the expression (\ref{eq:muI}) for the chemical
potential (again at the two--body correlation level):
\begin{eqnarray}
&&      {1\over2}\Delta\left\{
        \int d^3r_1 d^3r_2\,\rho(\rvec_1,\rvec_2)
        \left[ V(\vert\rvec_1-\rvec_2\vert)
        - {\hbar^2\over8m}(\naa1+\naa2)
        u_2(\rvec_1,\rvec_2)\right]\right\}
\nonumber\\
&&      =-{1\over2}\int d^3r_1 d^3r_2\,\rho(\rvec_1,\rvec_2)
         {\hbar^2\over8m}(\naa1+\naa2)
        \Delta u_2(\rvec_1,\rvec_2)\,.
\label{eq:muI1}
\end{eqnarray}
In particular, since $T_{\rm JF}$ can be expressed entirely in
terms of the $\Gamma_{\rm dd}(r_{ij})$ and exchange functions,
there is no rearrangement correction from this term.
The change, $\Delta u_2$, is now expressed as a functional of the
impurity quantities,
\begin{eqnarray}
        \Delta u_2(r) &&\equiv \Delta u_2[\rho^I,g^{\rm IB},u_2^{\rm IB}]
\nonumber\\
        &&=\int d^3r_0\,\rho^I\,{\delta u_2(\rvec_1,\rvec_2)
        \over\delta \rho^I}\,.
\end{eqnarray}
Furthermore using the HNC equation we find, again in HNC
approximation, that
\begin{equation}
\Delta u_2(\rvec_1,\rvec_2) =
        - \int d^3r_0\,\rho^I\,
        {\delta N_{\rm dd}(\rvec_1,\rvec_2)\over\delta \rho^I}\,.
\label{dubb}
\end{equation}
The variations are carried out for {\it fixed pair distribution
functions.\/} The change in the sum of nodal diagrams has a simple
expression in the momentum space,
\begin{equation}
        \rho^I{\delta \tilde N_{\rm dd}(k)\over\delta \rho^I}
        = \Bigl(\tilde X_{\rm dd}^{\rm IB}(k) \Bigr)^2.
\label{dnbb}
\end{equation}

The HNC equations (\ref{eq:Gddr}) and (\ref{eq:I_hypernet}) are now
used to eliminate the pair correlation functions $u_2(r)$ and
$u_2^{\rm IB}(r)$ from the chemical potential (\ref{eq:muI}). Using
the HNC equations and Eq. (\ref{dnbb}) one finds after some algebraic
manipulations
\begin{eqnarray}
        \mu^I_{HNC}&=& \rho\int d^3r
                \Biggl[ g^{\rm IB}(r)
        v^{\rm IB}(r)+\left[\hmI2 +\hm2\right]
        \left\vert{\bm \na}\sqrt{ g^{\rm IB}(r)}\right\vert^2\Biggr]
\nonumber\\
        &+&{1\over 4}
        \int {d^3k\over (2\pi)^3\rho} \left[t(k)+t_I(k)\right]\left[
        S^{\rm IB}(k)
        \left[\tilde N_{\rm dd}^{\rm IB}(k) 
        + \tilde E_{\rm dd}^{\rm IB}(k) \right]
        -\left[\tilde N_{\rm de}^{\rm IB}(k) 
        + \tilde E_{\rm de}^{\rm IB}(k) \right]
        \tilde \Gamma_{\rm dd}^{\rm IB}(k)\right]\nonumber\\
        &+&{1\over4}\int {d^3k\over (2\pi)^3\rho}\,
        t(k)\left[S(k)-1\right]
        \left[\tilde X_{\rm dd}^{\rm IB}(k)\right]^2\,.
\label{eq:muHNC}
\end{eqnarray}

The expression given above must be supplemented by corrections
originating from ``proper'' elementary diagrams and triplet
correlations (see Ref. \onlinecite{impu4in3}) if appropriate.

\section{Simplifications}
\label{ssec:fhnc00}

The FHNC--EL equations for the background, and even more for the
impurity, are admittedly complicated and of little appeal. The reason
is that the exchange structure allows for many different ways of
coupling.

We recall, however, that the original motivation for deriving these
equations is a symmetric treatment of short--and long-- ranged
correlations, and the stability criteria provided by the
optimization. One is therefore tempted to reduce the FHNC--EL
equations to a level that contains the {\it minimal\/} amount of
self--consistency and is, nevertheless, optimizeable. An appealing
feature of such a simplified theory is that the equations are hardly
more complicated than those of the Bose theory; they are also readily
applied to the much more demanding problem of inhomogeneous systems,
where the solution of the full set of FHNC--EL equations is a rather
unpleasant and still uncompleted task \cite{Surface4}. In this
connection, we need to recall two things: First, positron annihilation
rates are directly connected to short--ranged correlations, {\it
c.f.\/} Eq. (\ref{eq:tau}). Second, it is far from straightforward to
deduce short--ranged correlations in an inhomogeneous geometry from
results on the homogeneous electron gas \cite{KrKohnPRL}.

\subsection{FHNC//0 for the background}
\label{ssec:FHNCcc0}

The FHNC//0 approximation is, by construction, the {\it minimum\/}
approximation that is (a) correct for both short-- and long--range
correlations and (b) permits optimization. The FHNC//0 approximation
for the background amounts to ignoring all $de$ quantities and using
$1 + \tilde X_{\rm ee}(k) = S_F(k)$, where
\begin{equation}
        S_F(k)=\left\{
                \begin{array}{cc}
                \frac{3k}{4k_F}-\frac{k^3}{16 k_F^2}
            & k < 2 k_F     \\
                1    & k \geq 2 k_F
                \end{array}
        \right.
\end{equation}
is the static structure function of the non--interacting Fermi gas.

Consistent with this approximation, we leave out ``elementary''
diagrams; they may be put in at the end on a term--by--term basis. The
FHNC equations (\ref{eq:Gddr}), (\ref{eq:NodalF}) collapse to three
equations:

\begin{eqnarray}
X_{\rm dd}(r)   &=& \exp\left[u_2(r)+N_{\rm dd}(r) 
\right]
                -1 -N_{\rm dd}(r)\,,
 \nonumber\\
\tilde N_{\rm dd}(k)    &=& {\tilde X_{\rm dd}(k)
        \over 1 - S_F(k) \tilde X_{\rm dd}(k)}\label{eq:FHNC0}\,,\\
S(k) &=& {S_F(k)\over 1 - S_F(k) \tilde X_{\rm dd}(k)}\,.\nonumber
\end{eqnarray}
Eqs.~(\ref{eq:FHNC0}) are exact in the long--wavelength limit, but can
be applied also at finite momenta.  In the same approximation, we
obtain from Eq.~(\ref{eq:FermiEulerK})
\begin{eqnarray}
S(k) &=& {S_F(k)\over\sqrt{1 + 2 {S_F^2(k)\over t(k)}\tilde V_{\rm p-h}(k)}}
\nonumber\,,\\
        V_{\rm dd}(r) &=&  \left[1+ \gddr\right]v(r)
        + {\hbar^2\over m}\left|\nabla\sqrt{1+\gddr}\right|^2
+ \gddr w_{\rm I}(r)\nonumber\,,\\
        \tilde w_{\rm I}(k)&=&-{\hbar^2k^2\over 4m}
        \left[{1\over S_{\rm F}(k)}-{1\over S(k)}\right]^2
        \left[2{S(k)\over S_{\rm F}(k)}+1\right].
\end{eqnarray}
For further reference, we also mention that, in the same
approximation,
\begin{equation}
\tilde \Gamma_{\rm dd}'(k) = {1\over2}\,t(k)\tilde\Gamma_{\rm dd}(k)
\left[1-{2\over S_{\rm F}(k)}\right]\,.
\label{eq:Gddprime0}
\end{equation}

\subsection{FHNC//0 for the impurity}

Leaving the FHNC equations of section \ref{ssec:FHNC_I} unchanged but
identifying $1 + \tilde X_{\rm ee}(k) = S_F(k)$ and $S(k)/S_{\rm d}(k)
= S_F(k)$ and ignoring $\tilde V_{\rm de}(k)$ and $\tilde V_{\rm
ee}(k)$, the impurity Euler equations reduce to

\begin{equation}
\tilde {\cal V}^{\rm IB}(k) = \tilde V^{\rm IB}_{\rm dd}(k) +
{\tilde V^{\rm IB}_{\rm de}(k) \over S_F(k)} - 
{t(k)\over2}{\tilde X^{\rm IB}_{\rm de}(k)\over S_F^2(k)}
\left[S_F(k) -1\right]\,.
\end{equation}

The induced interaction is
\begin{eqnarray}
\tilde w^{\rm IB}_{\rm I}(k) &=&
-{1\over2}\left[t_I(k)+{t(k)\over S_F(k)}\right]
\tilde\Gamma^{\rm IB}_{\rm dd}(k)
 -{1\over S_F(k)}\left[
\tilde V^{\rm IB}_{\rm de}(k)
-{1\over2}\left[t_I(k)+t(k)\right]\tilde X^{\rm IB}_{\rm de}(k)\right]
\nonumber\\
&-& \tilde V^{\rm IB}_{\rm de}(k)\,.
\end{eqnarray}
The $de'$--equation does not simplify significantly, note especially
that it is not legitimate to omit the ``elementary'' exchange
diagrams.

Let us finally also ignore all impurity $de$ quantities.
The Euler equation (\ref{eq:eulerI}) remains
the same, we only need to replace ${\cal X}^{\rm IB} \rightarrow
X^{\rm IB}_{\rm dd}$
and ${\cal V}^{\rm IB}\rightarrow V^{\rm IB}_{\rm dd}$. We can write it also as
\begin{equation}
S^{\rm IB}(k) = -2 {S(k)\tilde V^{\rm IB}_{\rm dd}(k) \over t_I(k)
+ \hbar\omega(k) } \,,
\label{eq:XddIB0}
\end{equation}
\begin{equation}
V^{\rm IB}_{\rm dd}(r) = 
\frac{\hbar^2}{m}\left| \nabla \sqrt{1+\Gamma^{\rm IB}_{\rm dd}(r)}\right|^2 + 
\left( 1 + \Gamma^{\rm IB}_{\rm dd}(r) \right) \,v^{\rm IB}(r) +
\Gamma^{\rm IB}_{\rm dd}(r)\,w^{\rm IB}_{\rm I}(r)\,.
\end{equation}
The induced interaction is
\begin{eqnarray}
\tilde w^{\rm IB}_{\rm I}(k) &=&- \tilde V^{\rm IB}_{\rm dd}(k) - {1\over2}
\left[ {t(k) \over S_F(k)}
+ t_I(k)\right]\tilde \Gamma^{\rm IB}_{\rm dd}(k)\nonumber\\
&=& {1\over2}\,S^{\rm IB}(k)\,
\left( {t(k)\over S^2(k)}- {t(k)\over S_F^{2}(k)} +
\frac{t_I(k)}{S(k)} 
- \frac{t_I(k)}{S_F(k)} \right)
\end{eqnarray}
and finally
\begin{equation}
\tilde \Gamma^{\rm IB}_{\rm dd}(k) = \frac{S^{\rm IB}(k)}{S_F(k)}\,.
\label{eq:GddIB0}
\end{equation}

We can apply the above sequence of simplifications in two ways: One
way is to use the FHNC//0 approximation for the background only,
keeping all impurity equations intact. This might be a possible
strategy for calculating annihilation rates in inhomogeneous
geometries, the good agreement between the $g^{\rm IB}(0)$ values in
this approximation with the Monte Carlo data of Ortiz
\cite{OrtizThesis} provides encouragement that this is perhaps a
pragmatic way to attack this problem.

The second level of approximation is to also omit the $de$--diagrams
in the impurity equations. We note that {\it both\/} strategies
include the self--consistent summation of ring-- and ladder-- diagrams
\cite{parquet1,parquet2}, {\it both\/} strategies also maintain
``perfect screening'' $S^{\rm IB}(k) = 1$. The FHNC//0 approximation
provides for bulk electrons an accuracy that is only marginally worse
than the full theory \cite{Surface4}, and one might hope that the same
is true for electron--positron mixtures. However, since the above
approximation omits the self--consistent summation of ``elementary''
exchange diagrams, the pair distribution function and the structure
function become inconsistent. A simple approximation for $g^{\rm
IB}(r)$ would be
\begin{equation}
g^{\rm IB}(r) \approx 1 + \Gamma_{\rm dd}^{\rm IB}(r)\,.\nonumber
\end{equation}
Adding the simplest exchange term would lead to a different
approximation,
\begin{equation}
g^{\rm IB}(r) \approx \left[1 + \Gamma_{\rm dd}^{\rm IB}(r)\right]
\left[1 + C(r)\right]\,,\nonumber
\end{equation}
where $\tilde C(k) = \tilde \Gamma_{\rm dd}^{\rm IB}(k)
\left[S_F(k)-1\right]$. This is acceptable in a system with repulsive
interactions since $1 + \Gamma_{\rm dd}^{\rm IB}(r)$ is small at the
origin. However, the opposite is true for the electron--positron case
because both $g^{\rm IB}(r)$ and $1 + \Gamma_{\rm dd}^{\rm IB}(r)$
become large at the origin, thus enhancing any small error that might
be made in calculating the exchange corrections. We will see in the
next section that this effect can be quite dramatic.

\section{Annihilation rates}

The central quantity of interest of our calculation is the positron
annihilation rate (\ref{eq:tau}), hence, our primary question is {\it
what does it take to get $\tau$ right ?\/} To examine this question,
we have carried out the following sequence of calculations of
increasing complexity:
\begin{itemize}

\item Charged bosons:  Positive impurity in a charged Bose gas.

\item FHNC//0 : The basic version of the FHNC--EL theory for both the
background and the impurity. This means we use Eqs.
(\ref{eq:FHNC0})--(\ref{eq:Gddprime0}) for the background, and Eqs.
(\ref{eq:XddIB0})--(\ref{eq:GddIB0}) for the impurity. The motivation
for this approximation is that it is the simplest one that contains
fermionic corrections, satisfies all exact short-- and
long--wavelength properties, and can therefore be optimized.

\item FHNC//0b: The full impurity FHNC--EL equations on the
simplified background.  The motivation for this approximation is that
there is evidence that a simplified treatment of the background
electrons could be appropriate, in fact, the energetics of the bulk
electron gas predicted by this approximation is not significantly
worse than the energetics produced by the full theory
\cite{Surface4}. However, electrons and positrons are strongly
correlated and, thus, the same approximation might not be adequate for
electron--positron correlations. Since the treatment of the impurity
correlations is identical to the one of the full FHNC--EL theory to be
described below, the comparison of the results from this calculation
with the ones from the next more sophisticated one examines the
sensitivity of the results to the background correlations.

\item FHNC//C0 : The full solution of the FHNC--EL equations
for the electrons and the positronic impurity, including corrections
that guarantee the long--wavelength limit (\ref{eq:XIBdelong}) but
{\it without\/} proper elementary diagrams.

\item FHNC//C5 : As FHNC//C0, but including 4$^{\rm th}$
and 5$^{\rm th}$ order proper elementary diagrams as described in
Ref. \onlinecite{polish}.
\end{itemize}

Table \ref{tab:gib0} lists a comparison of our results for the primary
quantity, $g^{\rm IB}(0)$, in the simplest charged boson calculation
(see also Ref. \onlinecite{MikkoMBVIII}), our most sophisticated
FHNC//C5 calculation, and a few important earlier calculations
\cite{Kahana63,BorSzoSta81,Lantto87,StaL93,OrtizThesis}.  At high
densities, one has reasonable agreement whereas the agreement becomes
worse at lower densities. The comparison of different methods should
serve to identify the source of these discrepancies. Among the various
calculations, the variational Monte Carlo calculations of Ortiz
\cite{OrtizThesis} have produced consistently the lowest values for
$g^{\rm IB}(0)$; incidentally, these values are quite close to
our FHNC//0b results.

\begin{table}
\begin{center}
\begin{tabular}{l |ddddddd|}
$r_s$ & \multicolumn{1}{c}{~Kahana} & \multicolumn{1}{c}{~~BSS~~} & 
 \multicolumn{1}{c}{~Lantto} & \multicolumn{1}{c}{~~~SL~~} &
 \multicolumn{1}{c}{~~Ortiz} & \multicolumn{1}{c}{~Bosons} & \multicolumn{1}{c}{FHNC//C5} \\
\hline
\hline
1  &       &       &  2.16 &  2.29  &  2.06 &   2.40  &   2.076   \\
2  & 2.21  &  3.76 &  4.05 &  3.96  &  3.39 &   4.55  &   3.983   \\
3  & 2.67  &  6.91 &  7.40 &  7.29  &  7.00 &   8.25  &   7.658   \\   
4  & 3.17  & 13.6  & 13.2  & 13.6   & 11.70 &  14.71  &  14.455  \\
5  &       &       & 23.0  & 24.2   & 17.69 &  26.08  &  26.225   \\
6  &       &       &       & 40.1   &       &  46.18  &  45.640   \\
8  &       &       &       & 93.9   & 64.00 & 143.30  & 126.063   \\
\hline
\end{tabular}
\end{center}
\caption{The table compares the primary quantity of interest, $g^{\rm IB}(0)$, 
for the ``charged boson'' approximation (Bosons) and the
FHNC//C5 approximation, with the values obtained by Kahana
\protect\cite{Kahana63}, Boro{\'n}ski, Szotek and Stachowiak (BSS)
\protect\cite{BorSzoSta81}, Lantto \protect\cite{Lantto87}, Stachowiak
and Lach (SL) \protect\cite{StaL93}, and Ortiz
\protect\cite{OrtizThesis}.}
\label{tab:gib0}
\end{table}

\begin{figure}
\includegraphics[width=0.8\linewidth] {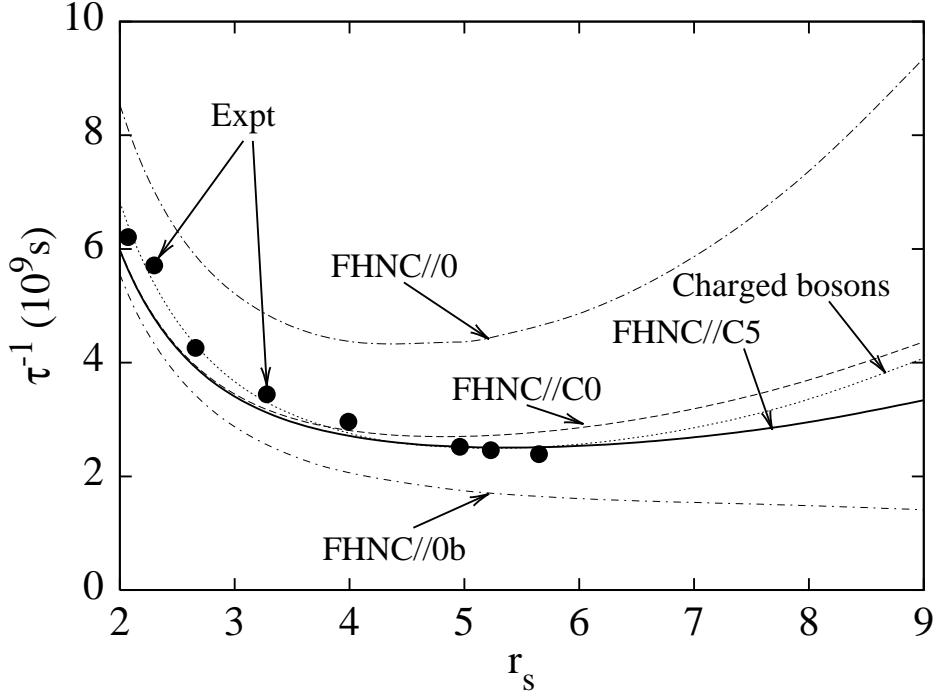}
\caption{Experimental and theoretical annihilation rates for various
materials. The experimental points are, from left to right, for the
metals Al, Zn, Mg, Li, Na, K, Rb, and Cs
\protect\cite{expSeeger,expHall,expWeisberg}, without core
corrections. The theoretical curves correspond to various levels of
implementation of the (F)HNC--EL theory as labeled in the figure. In
the level of sophistication the order from simple to complex is:
Charged bosons, FHNC//0, FHNC//0b, FHNC//C0, FHNC//C5 (see main text
for definitions). The positronium decay rate is (500~ps)$^{-1} =
2\times 10^9 s^{-1}$.  }
\label{fig:Annihilationrates}
\end{figure}

Fig.~\ref{fig:Annihilationrates} shows the annihilation rates
$\tau^{-1}$ computed using the approximations listed above.  More
appropriately, one should refer to these numbers as ``partial
annihilation rates'' because we have neglected the effect of core
electrons. Sob \cite{Sob85} obtains, for example, 5\% core
contribution for Li, and 30 \% for Cs, whereas more recent work
\cite{DSR91,ABH96} arrives at even larger core corrections. We
conclude first that, in the experimentally accessible density range,
the ``charged boson'' approximation is not as bad as one might
expect. Inclusion of the simplest fermion corrections (FHNC//0)
yields, on the other hand, far too high annihilation rates. This
failure is rather surprising, because the approximation already
includes the self--consistent summation of both ring and ladder
diagrams \cite{parquet1,parquet2}, and is thus beyond standard
perturbative treatments that include only one of the classes of
diagrams or an incomplete self--consistency scheme.

The main reason for the failure of the FHNC//0 approximation can be
traced to the way exchange corrections were included (see the
discussion in the end of section \ref{ssec:fhnc00}). On the level
FHNC//0b this problem is solved by including the full FHNC summation
for the impurity correlations, while leaving the background
correlations at the FHNC//0 level. Indeed, there is some improvement.
The annihilation rates come out lower than those predicted by the more
sophisticated approaches to be discussed below. But, in fact, when
core corrections are included, agreement of the FHNC//0b approximation
with experiments is not too bad. The full FHNC summation FHNC//C0
brings us back up, and ultimately the inclusion of ``proper''
elementary diagrams on level FHNC//C5 produces reasonable agreement
with the experimental annihilation rates when core corrections are
omitted.

Evidently $g^{\rm IB}(0)$ is extremely sensitive to the description of
the background electron liquid, as can be seen in the large difference
between the full FHNC//0, the FHNC//0b and the FHNC//C0 results. We
conclude therefore that any simplistic treatment of background
correlations should be viewed with much reservation.

In discussing the possible theoretical refinements one should finally
comment on the importance of triplet correlations. Since the proper
elementary diagrams have an impact, one might expect the same to be
true for triplets as well. In the case of helium liquids, triplets
have been mostly described in the ``convolution approximation''
outlined in Ref. \onlinecite{MixMonster}. Recently, the treatment has
also been found to be adequate for a problem analogous to the present
one, namely the calculation of the energetics of a \he4 impurity in
\he3 \cite{impu4in3}. The next systematic improvement to the method of
Ref. \onlinecite{MixMonster} has also been examined \cite{EKthree},
but for helium liquids it was found to be
insignificant. Unfortunately, the same is not true for positronic
impurities in the electron gas. We have tried the same approach here
and found that the ``convolution approximation'' is manifestly
inadequate for this case.  It appears that the derivation of triplet
HNC equations \cite{Wertheim} is necessary for a satisfactory
treatment. We are not aware of a successful implementation of this
program for bosons, much less for fermions or mixtures, and must leave
this question open for the time being. The variational Monte Carlo
calculations of Ortiz \cite{OrtizThesis} support our view that triplet
correlations are unimportant.

\begin{figure}
\includegraphics[width=0.8\linewidth]{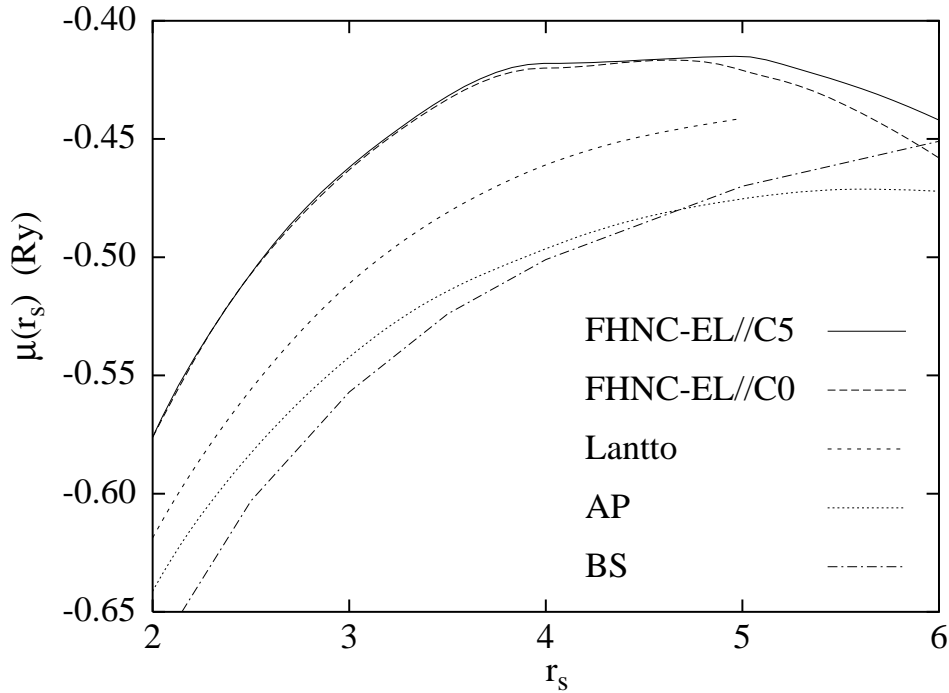}
\caption{The calculated positron correlation energies as function of
$r_s$ in the FHNC//C0 (dashed line) and the FHNC//C5 (solid line)
approximation. Also shown are the theoretical results of Lantto
\protect\cite{Lantto87}, Arponen and Pajanne (AP)
\protect\cite{anniArponen}, and Boro{\'n}ski and Stachowiak (BS)
\protect\cite{BorSt98}.}
\label{fig:mu}
\end{figure}

From now on we show only the FHNC//C5 result, and occasionally the
FHNC//0 result for reference. The next quantity of interest is the
positron correlation energy (or chemical potential)
(\ref{eq:chemi}). Our results are shown in Fig. \ref{fig:mu}; they are
consistent with earlier calculations of Arponen and Pajanne
\cite{anniArponen} and Lantto \cite{Lantto87}, although our results
are somewhat higher. The present results also show a drop of the
chemical potential towards lower densities; a similar drop has been
reported by Hodges and Stott \cite{HodgesStott}.

It has been argued that, in the low--density limit (large $r_s$), the
correlation energy should go towards the binding energy of the free
positronium of -0.524 eV. Evidently, our calculation does not show
such an asymptotic behavior and we hasten to point out that it should
not: The Jastrow--Feenberg function used in
Eqs.~(\ref{eq:JastrowWaveFunction}) and
(\ref{eq:ImpurityWaveFunction}) describes a state in which the
positron is delocalized, so it is {\it not\/} a valid description of
one positronium atom and $N$ free electrons. Hence, the present theory
is inapplicable to a system with bound electron--positron pairs. It
has, however, the desirable feature that the Euler equation ceases to
have a solution at the point where system cannot be in the state
described by the trial wave function. In the present case, this
happens at the density where the positron {\em must} be localized.
This feature of the theory does not preclude that the free positronium
is the physical low--density limit; the theory simply makes no
statement about this limit.

\begin{figure}
\includegraphics[width=0.8\linewidth]{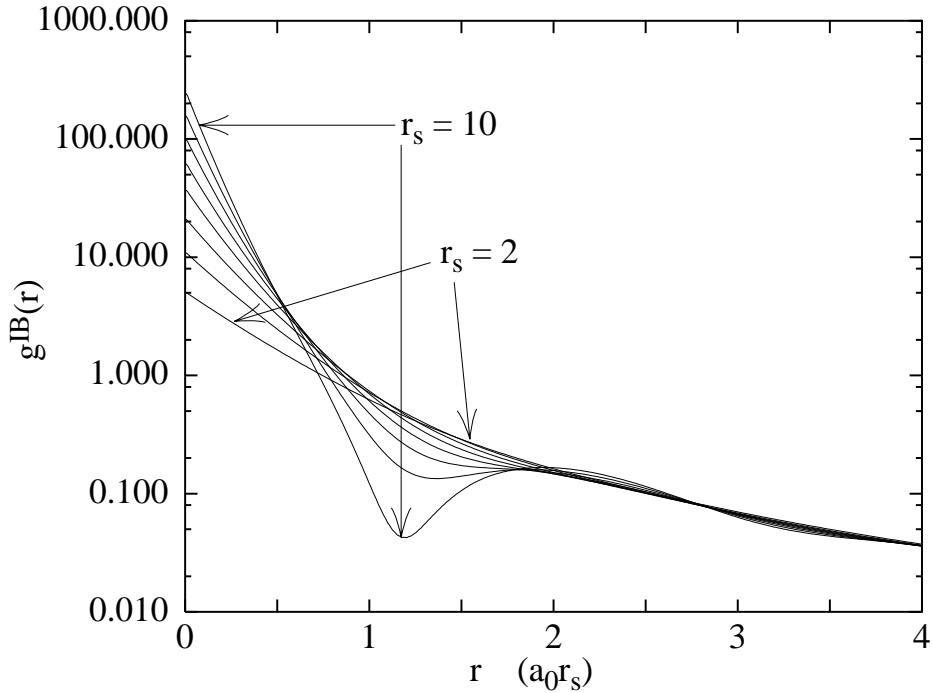}
\caption{The figure shows our calculated electron--positron pair
distribution function $g^{\rm IB}(r)$ for $r_s = 1, 2, \ldots, 9$.
The function with the lowest value of $g^{\rm IB}(0)$ corresponds to
the highest electron density. Notice the logarithmic scale in $g^{\rm
IB}(r)$.  }
\label{fig:gib}
\end{figure}

The appearance of a positronium atom is, on one hand, signaled by a
very large $g^{\rm IB}(0)$. Moreover, recall that we solve the Euler
equation with the boundary condition $g^{\rm IB}(r)\rightarrow 1$, as
$r \rightarrow \infty$. If there were a bound state, then
$\sqrt{g^{\rm IB}(r)}$ should have a node at some finite distance. Of
course --- as argued above --- the theory will not allow us to reach
this point, but one should expect precursor phenomena as one
approaches the density of positronium formation.  Fig. \ref{fig:gib}
shows indeed both effects: As we decrease the density, the $g^{\rm
IB}(0)$ increases. In addition to this, a ``dip'' forms at an
intermediate distance. At the highest $r_s$ value shown in
Fig. \ref{fig:gib}, $r_s = 9$, we have $g^{\rm IB}(r)$ as low as
0.042. With decreasing density the dip drops rapidly, and finally the
pair distribution function becomes unphysically negative at
$r_s\approx 9.4$.  Also $g^{\rm IB}(0)$ diverges, at $r_s = 10$, but
because of both diagrammatic approximations and numerical limitations
one cannot expect these instabilities to show up at exactly the same
density. We conclude that our calculations predict positronium
formation at $9.4\le r_s\le 10$. For comparison, in the mid 70's Lowy
and Jackson \cite{Lowy75} found the positronium limit at $r_s \geq
6.2$, while a decade later Pietil\"ainen and Kallio \cite{PitKal83}
obtained $r_s \geq 8$.

\begin{figure}
\includegraphics[width=0.8\linewidth]{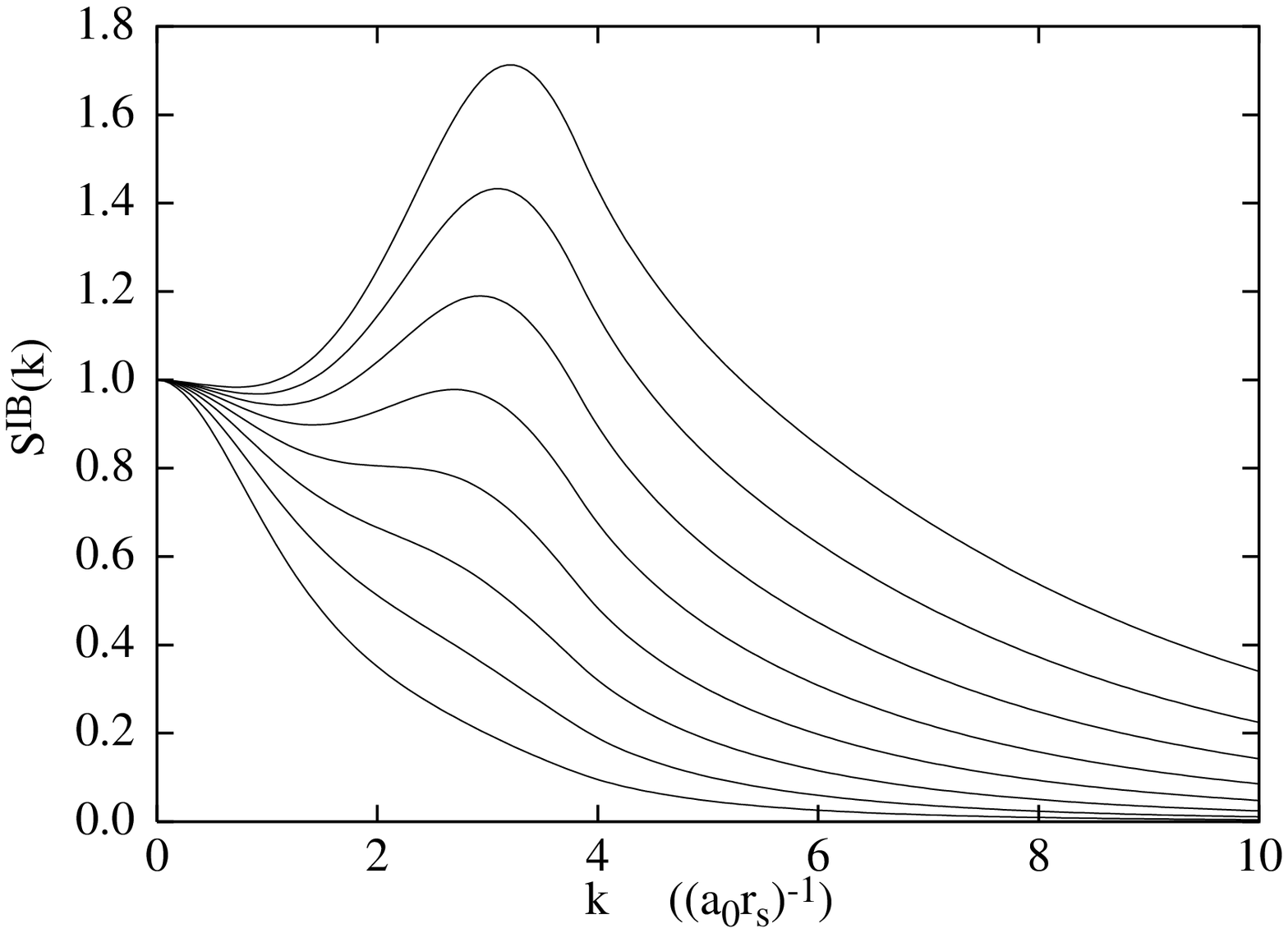}
\caption{The figure shows our calculated electron--positron structure
function $S^{\rm IB}(k)$ for $r_s = 2, 3, \ldots, 9$.  The lowest
curve corresponds to the lowest $r_s$ (highest electron density).}
\label{fig:sib}
\end{figure}

Slightly different information is contained in the static
electron--positron structure function $S^{\rm IB}(k)$, depicted in
Fig.~\ref{fig:sib}. In the long wavelength limit, charge conservation
implies $S^{\rm IB}(0+) = 1$; this property is a rigorous feature of
the optimized structure function in any level of the FHNC--EL
approximations. As the density decreases, we see that $S^{\rm IB}(k)$
develops a peak at $k \approx 3.2 (a_0 r_s)^{-1}$. Such a peak
reflects an oscillatory structure in the pair distribution function.

\section{Summary}

We have presented in this work calculations of electron--positron
correlations in simple metals. Technically, we have used the most
complete summation of diagrams within the Jastrow--Feenberg theory.
We have shown that such a highly summed theory is indeed necessary for
a reasonably reliable prediction of the relevant quantities.  The
omission of triplet correlations is an unsatisfactory point, but
progress towards a fully consistent summation of three--body (F)HNC
equations and the corresponding Euler equation is not in sight. In any
event, we have demonstrated that one might have to do even more, but
one must definitely not do less than was done in our work to obtain a
conclusive answer.

Our results are, from a pragmatic point of view, somewhat
disappointing in the sense that the system is evidently much more
complicated than the bulk electron gas at metallic densities.  Bulk
electrons at metallic densities are one of the simplest systems to be
treated within microscopic many--body theory, and simple approximation
provide already reasonable agreement with exact results. The reason
for this problem is evidently the large value of the
pair--distribution function at the origin, which causes poor
convergence of diagrammatic summations. On the other hand, the
FHNC//0b approximation seems to provide reasonable agreement with
experiments when core--corrections are included.  With sufficient
caution, since better agreement with experiments does not indicate a
better theoretical treatment, one might be able to use this version of
the theory in a non--uniform environment and obtain reasonable
annihilation rates for positrons in simple metal surfaces \cite{Surface4}.


\section*{Acknowledgments}
This work was supported, in part, by the Austrian Science Fund under
project P12832--TPH. Extensive discussions and communications with
H. Sormann and the communication of unpublished results by G. Ortiz
are gratefully acknowledged

\newpage
\bibliography {papers}
\end{document}